\documentclass[slac_one]{revtex4}
\usepackage{graphicx}
\usepackage{fancyhdr}
\newcommand{\mygev}{\;\rm{Ge}\kern-0.06667em\rm{V\/}}
\newcommand{\mytev}{\;\rm{Te}\kern-0.06667em\rm{V\/}}
\newcommand{\mypt}{p_{\rm T}} 
\newcommand{\myet}{{E}_{\rm T}} 
\def\myetmiss{\mbox{\ensuremath{\, \slash\kern-.6emE_{T}}}}

\newcommand{\mypb}{\rm{pb}}
\newcommand{\mycms}{\rm{cm^{-2}}\rm{s}^{-1}}

\newcommand{\FB}{$\rm\ fb^{-1}$}
\newcommand{\ttbar}{$t\bar{t}$}

\begin{document}

\title{Single-top Cross Section Measurements at ATLAS} 

%

\author{P. Ryan on behalf of the ATLAS Collaboration}
\affiliation{Michigan State University, East Lansing, MI 48824, USA}

%
%
\begin{abstract}
The single-top production cross section is one third that of the top-pair
production cross section at the LHC.  
During a year of data-taking, assuming an average luminosity
of $10^{33} \mycms$ and a CMS energy of $14 \mytev$, 
the determination of the major contributions to the total single-top cross
section should be achievable.
Comparisons between the measured cross sections
and the theoretical predictions will provide a crucial test of the
standard model. These measurements should also lead to the 
first direct measurement of $|V_{tb} |$, with a precision at the level
of a few percent.  In addition, they will probe for new physics via 
the search for evidence of anomalous couplings to the top quark and 
the measurements of additional bosonic contributions to single-top production.
Methods developed to optimize the selection of single-top
events in the three production channels are presented and the
potential for the cross section measurements with 1\FB\ and 30\FB\ of
integrated luminosity is established.
\end{abstract}

\maketitle

\thispagestyle{fancy}


%
%
\section{Single-Top Physics}
\begin{figure}
  \begin{center}
    \begin{tabular}{ccc}
      
      \begin{minipage}{5.1cm}
        \centering
        \includegraphics[height=3cm,width=3cm]{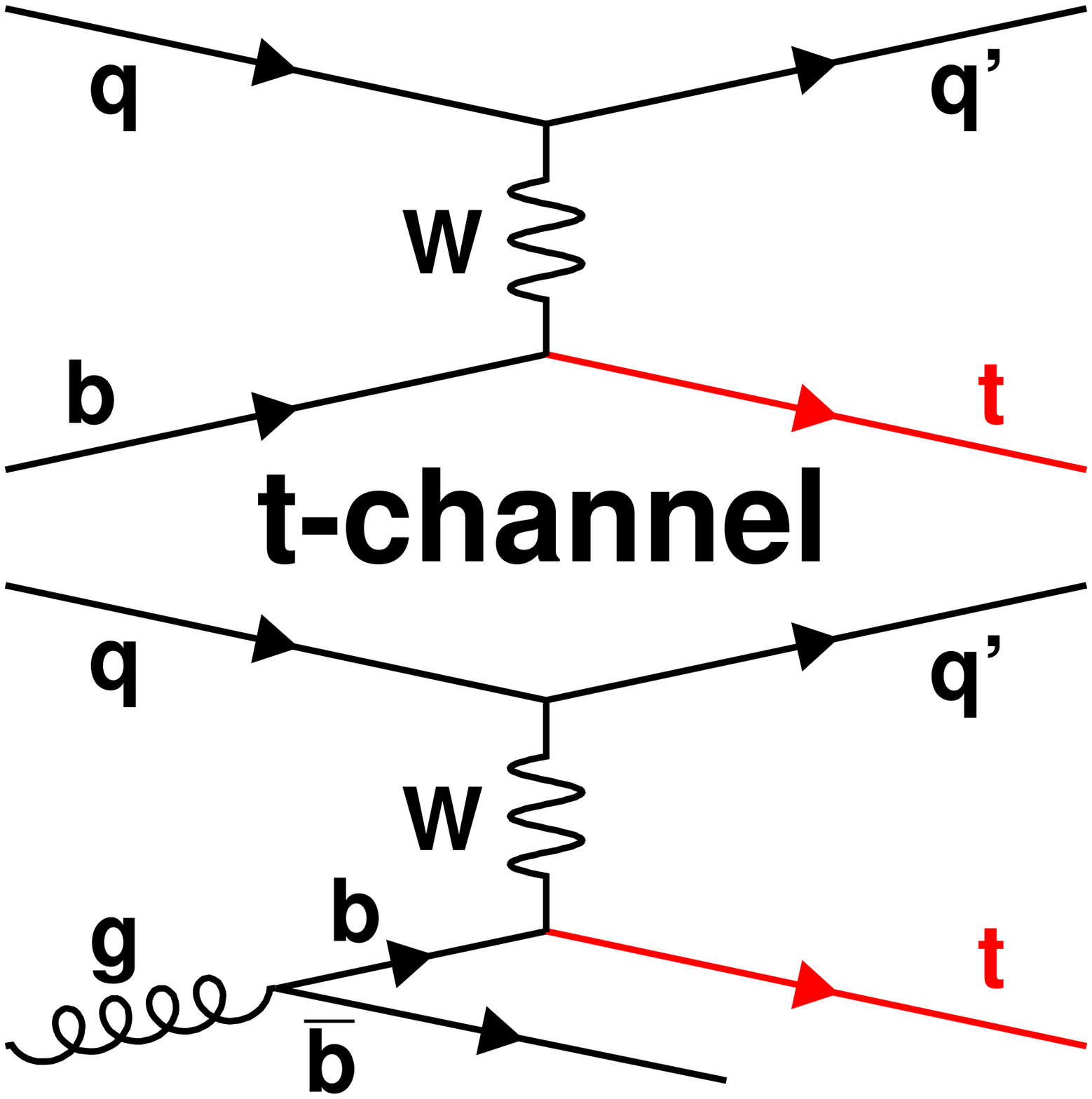}
      \end{minipage}
      
      \begin{minipage}{5.1cm}
        \centering
        \includegraphics[height=3cm,width=3cm]{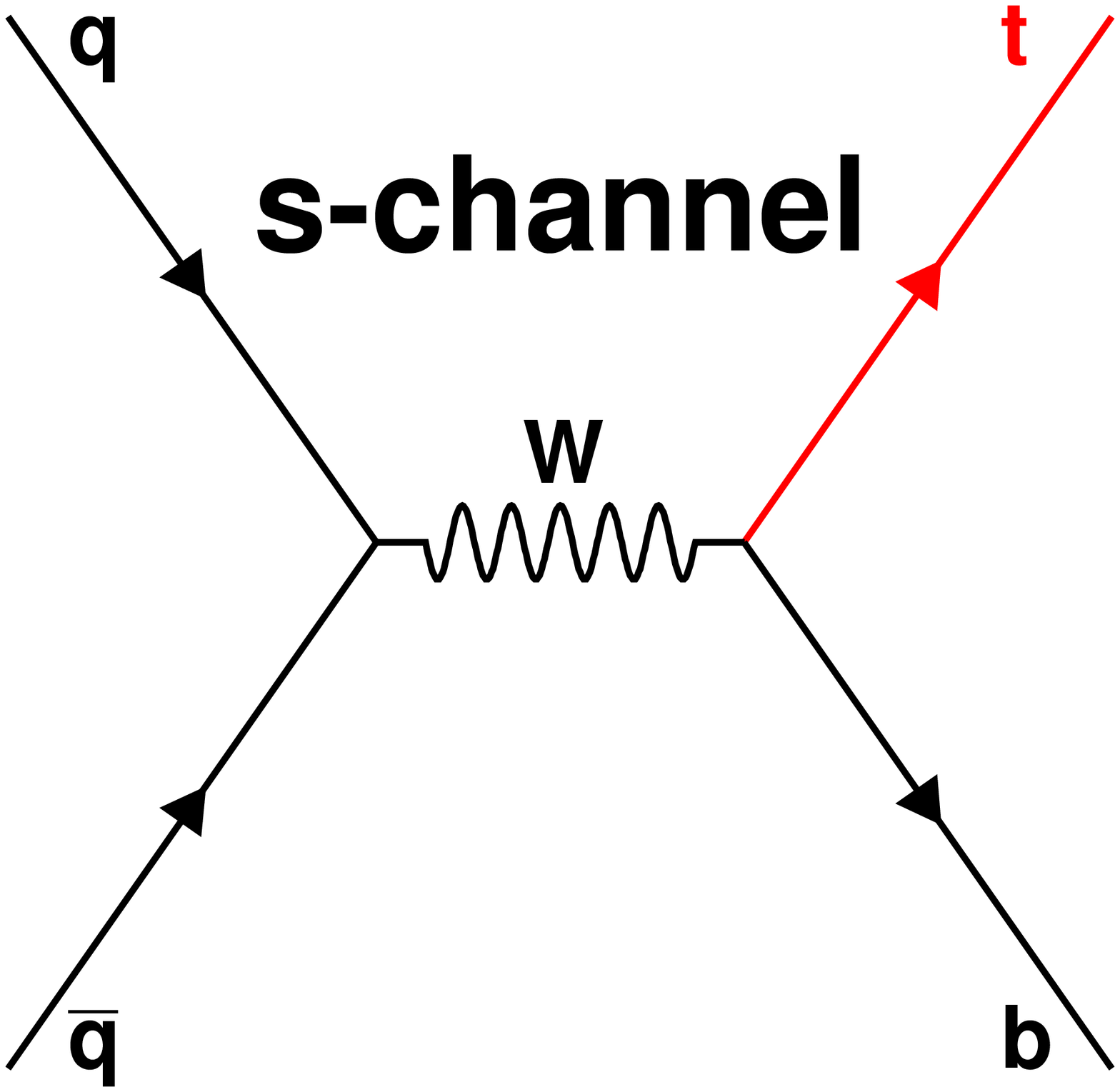}
      \end{minipage}

      \begin{minipage}{5.1cm}
        \centering
        \includegraphics[height=3cm,width=3cm]{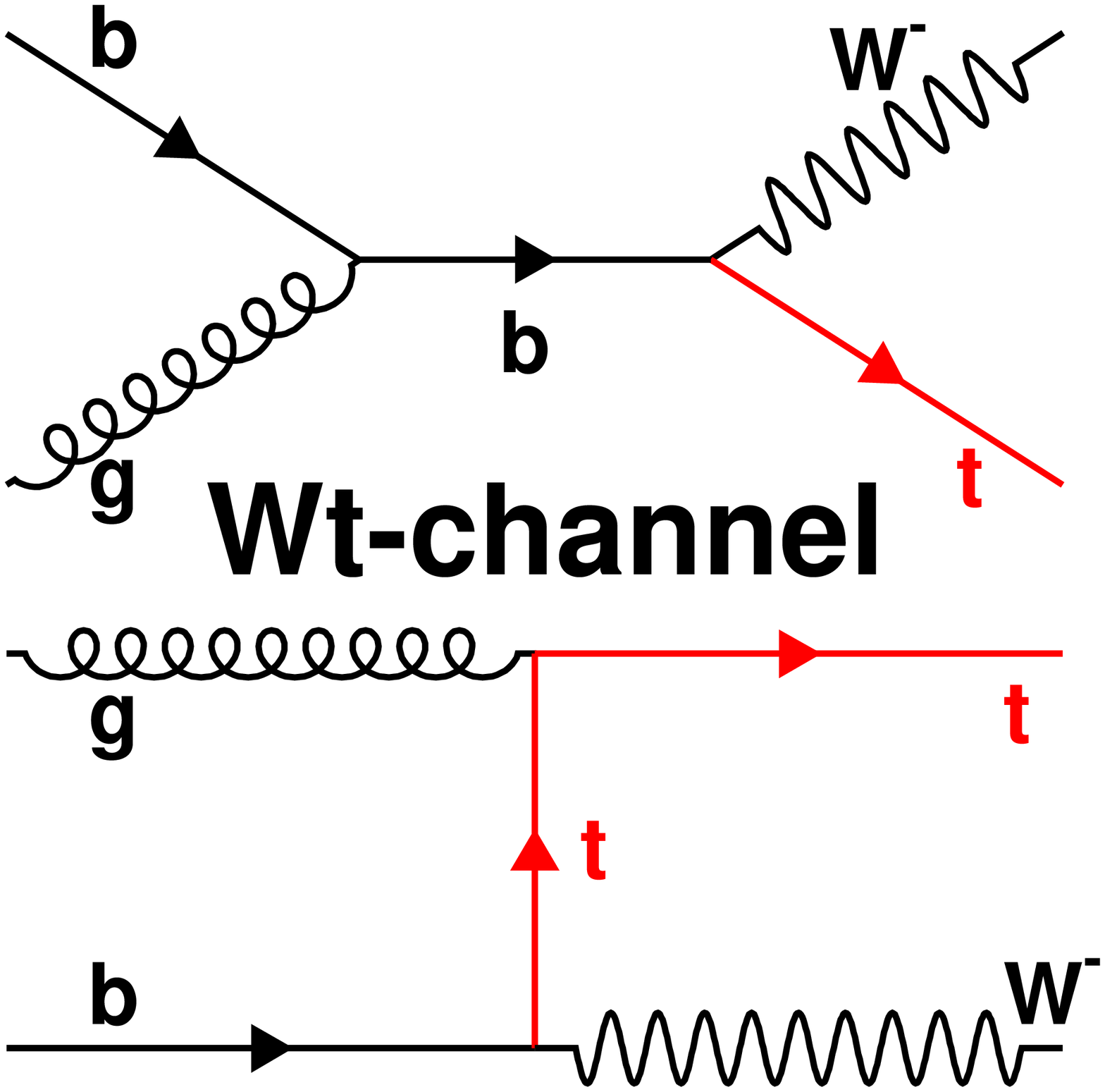}
      \end{minipage}
      
    \end{tabular}

    \caption[Single-top production diagrams.]
            {The diagram on the left shows t-channel single-top production, the diagram in 
              the middle shows s-channel single-top production, and 
              the diagram on the right shows $Wt$-channel single-top production.}
    \label{fig:singletop}
  \end{center}
\end{figure}

Single-top quarks are produced via the electroweak interaction and at leading order there are three
different production mechanisms; s-channel, t-channel, and $Wt$-channel, each of which are depicted in
Figure~\ref{fig:singletop}. 
The theoretical cross sections for the single-top processes are $246 \pm 10.2 \mypb$~\cite{Sullivan,Campbell} for the t-channel,
$10.65 \pm 0.65 \mypb$~\cite{Sullivan,Campbell} for the s-channel, 
and $66.5 \pm 3.0 \mypb$~\cite{Campbell2} for the $Wt$-channel.
Note that each single-top process has a $W-t-b$ vertex.
Only single-top events with an isolated and high-$\mypt$ electron or muon in the final state
are included in this study since events with only hadrons in the final state are difficult to
distinguish from background.  

The dominant background to single-top production is \ttbar\ production, 
which has a cross section three times larger than the combined single-top production cross section.
With a single high-$\mypt$ lepton, two $b$-jets, and $\myetmiss$, semi-leptonic \ttbar\ decay is most likely to mimic 
single-top events.
Due to the
lack of leptons in the final state, fully-hadronic \ttbar\ events are not a major contributor
to single-top background.
$W$ + jets production constitutes a significant background
since the cross-sections for these processes are several orders of magnitude greater than the
single-top cross sections.  
The background from di-boson events is minimal.
QCD will be estimated by data driven methods and is not considered in these studies.  
The amount of QCD contamination depends on the specific selections used in the analyses.

The single-top processes were generated using PYTHIA~\cite{PYTHIA}, with the matrix element calculated using AcerMC~\cite{ACERMC}.
The \ttbar\ sample, which included both di-lepton and semi-leptonic decays, was generated using HERWIG~\cite{HERWIG} and the matrix
element was calculated using MC@NLO~\cite{MCNLO}.  
The $W$ + jets sample was generated using PYTHIA and the matrix element was
calculated with ALPGEN~\cite{ALPGEN}.  MCFM~\cite{MCFM} was used to derive the K-factors needed to scale the LO processes
calculated with ALPGEN to NLO.

%
%
\section{Single-Top Pre-selection}
The three single-top processes shared a common pre-selection.  Muons and electrons were required to satisfy
reconstruction requirements of $\myet > 10\mygev$ and $|\eta| < 2.5$ and an isolation requirement
of $\myet < 6\mygev$ in a cone of radius 0.2 around the particle axis.
Events were required to contain one muon or electron with $\mypt > 30\mygev$ and events with secondary leptons
were removed to eliminate contamination from di-lepton \ttbar\ events and to ensure the orthogonality 
of the muon and electron samples.
Jet candidates were reconstructed using a cone algorithm with $\Delta R = 0.4$ and were required to satisfy
$\mypt > 15\mygev$ to be considered a jet.  An event was required to have between 2 and 4 jets, with
at least two of the jets having $\mypt > 30\mygev$ and at least one of the jets being $b$-tagged.  
Events were required to have $\myetmiss > 25 \mygev$,
which corresponds to the energy of the non-detected neutrino present in leptonic $W$ decay.

%
%
\section{Cross Section Measurements}
The measurement of the single-top cross sections in the ATLAS detector~\cite{ATLAS-Detector}
will be obtained using the formula
$$\sigma = \frac{N_{Data} - N_{bkg}}{\epsilon_{S} \times \cal{L}},$$
where $N_{Data}$ is the total number of events in the data, $N_{bkg}$ is the
number of expected background events, $\epsilon_{S}$ is the selection efficiency
for single-top signal events, and $\cal{L}$ is the luminosity.
Cross section errors were estimated by randomly generating
$N_{Data}$ according to a Poisson distribution and randomly varying $N_{bkg}$ and
$\epsilon_{S}$ for every systematic quantity 
by an amount determined by a Gaussian distribution around the central value of that quantity.
The sources of experimental uncertainty were
Jet Energy Scale (JES), $b$-tagging likelihood, and luminosity.
The sources of theoretical uncertainties were background cross sections, 
Initial State Radiation (ISR) and Final State Radiation (FSR),
PDFs, and $b$ quark fragmentation.

Cut-based and multivariate analyses of the cross section measurements 
were performed for each of the three single-top channels in order to
understand the size of the statistical and systematic errors and the amount
of integrated luminosity needed to obtain evidence and achieve discovery of the 
single-top quark.  A full description of the single-top analyses can be found in~\cite{CSC-Note}.

\begin{figure}
  \begin{center}
    \begin{tabular}{ccc}
      
      \begin{minipage}{7.5cm}
        \centering
        \includegraphics[height=4.5cm,width=7.37cm]{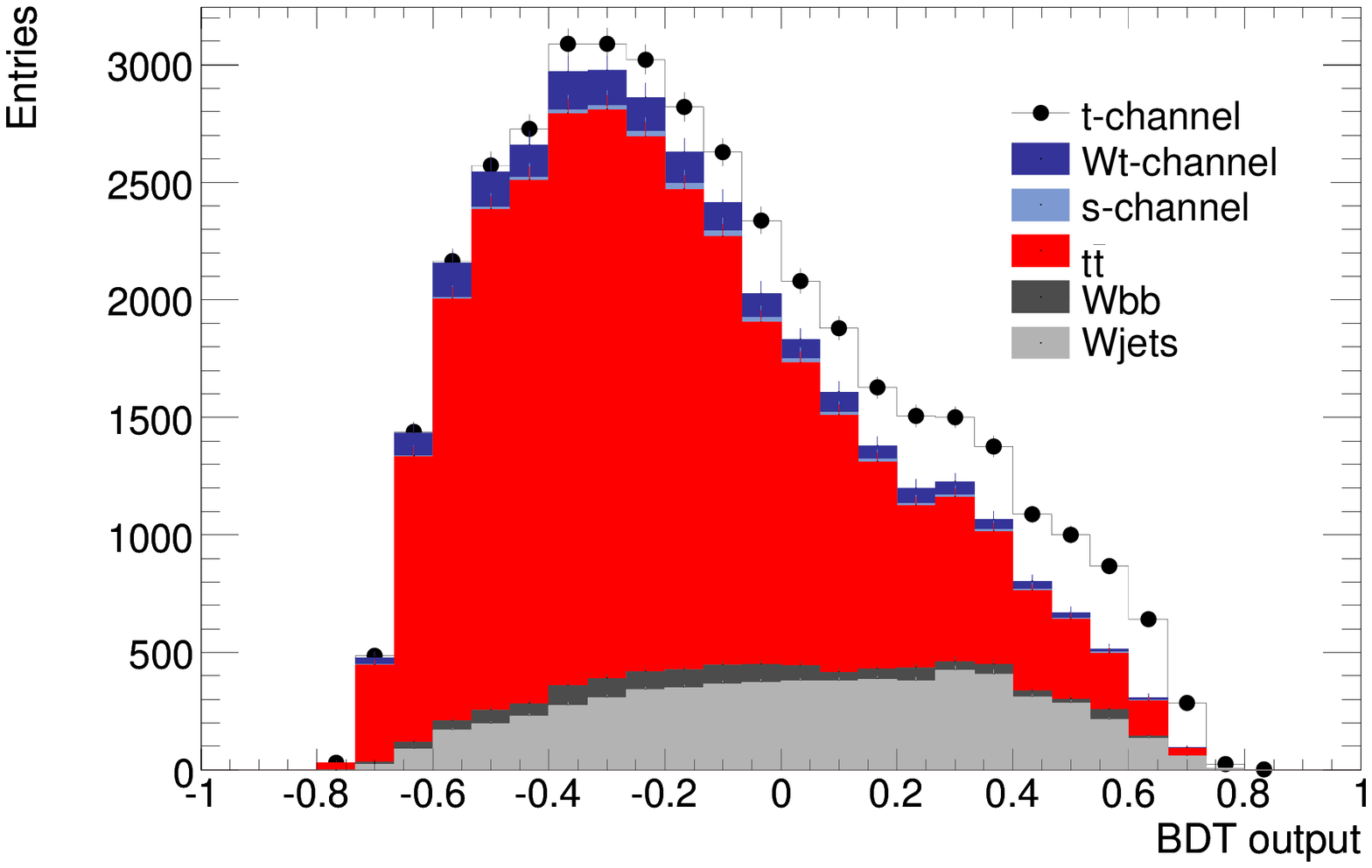}
      \end{minipage}
      
      \begin{minipage}{5.1cm}
        \centering
        \includegraphics[height=4.5cm,width=4.89cm]{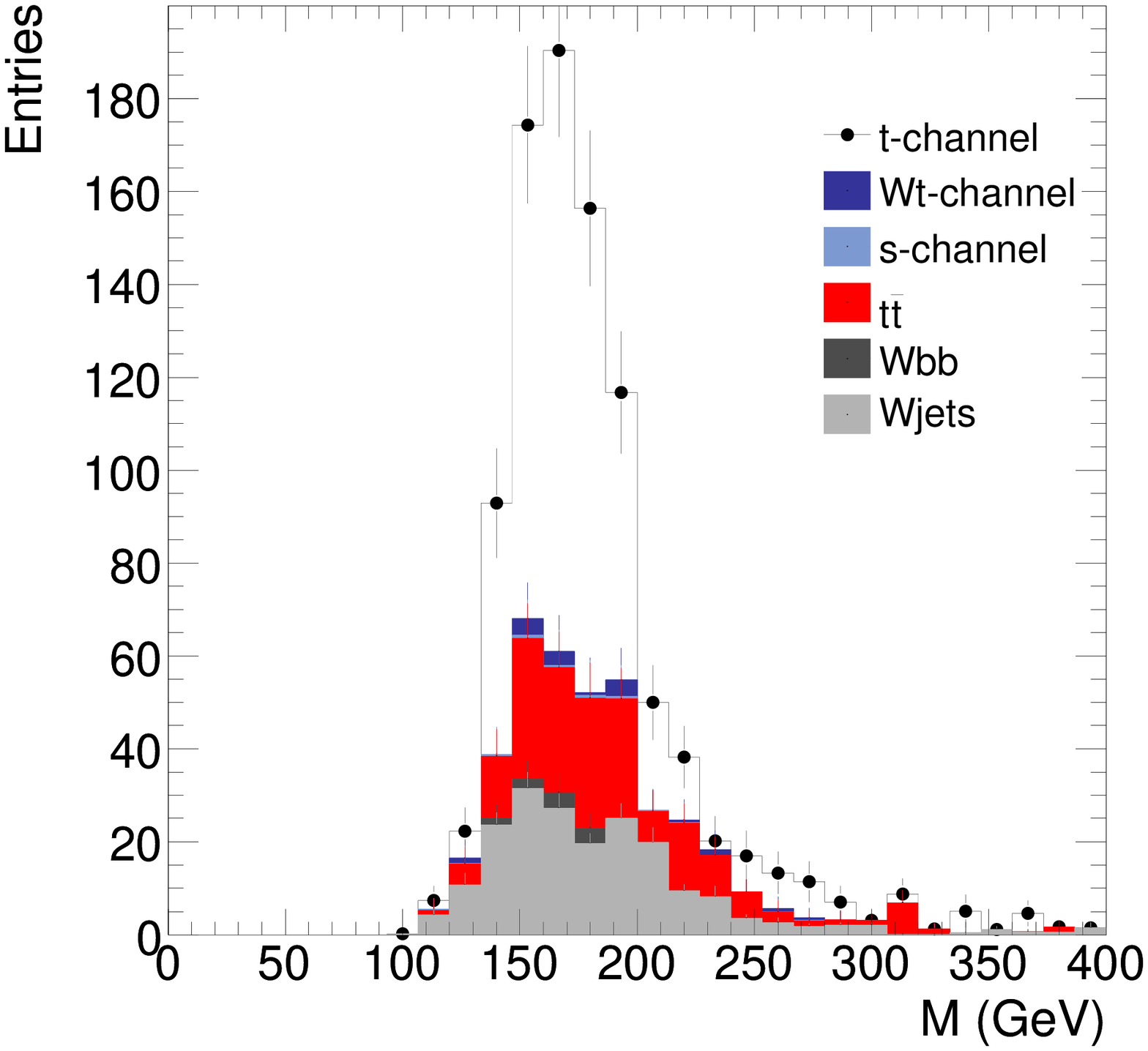}
      \end{minipage}

    \end{tabular}

    \caption[t-channel BDT output and the leptonic top mass distribution using a BDT cut of 0.6.]
            {The diagram on the left shows the BDT output for signal and background and the
	      diagram on the right shows the leptonic top mass distribution using a cut
	      of 0.6 on BDT output.}
    \label{fig:t-channel:BDT}
  \end{center}
\end{figure}

The cut-based analysis for the t-channel was performed by requiring, in addition to the
event pre-selection, a $b$-tagged jet with $\mypt > 50\mygev$ in order to remove
low-$\mypt$ $W$ + jets background and $|\eta| > 2.5$ for the hardest light
jet to remove \ttbar\ contamination.  After this selection, there
were 1,460 signal and 3,906 background events for a 1\FB\ sample.
The quantity of \ttbar\ background remaining after the
t-channel specific event selection
necessitated the use of a Boosted Decision Tree (BDT).
Variables providing a good
signal to background separation were used as input to the BDT and a cut of 0.6 on the BDT output
minimized the total uncertainty on the cross section and corresponded 
to a signal over background ratio of 1.3.
The BDT output is shown on the left side of Figure~\ref{fig:t-channel:BDT} and
the leptonic top mass distribution using a cut of 0.6 on the BDT output
is shown on the right side of Figure~\ref{fig:t-channel:BDT}.
%
Table~\ref{tbl:t-channel:errors} lists the uncertainties for 1\FB\ and 10\FB\ of integrated luminosity.
For both analysis methods and both integrated luminosity samples, the
systematic uncertainties far outweigh the statistical uncertainties.
The main systematic uncertainties on the t-channel cross section measurement
were JES, ISR and FSR, and luminosity.  
The systematics achieved using the BDT were approximately half of those obtained with the cut based analysis.  
\begin{table}[ht]
  \begin{center}
    \begin{tabular}{|c|c|c|c|c|}
      
      \hline
      {\bf Selection} & {\bf Luminosity} & {\bf $\Delta\sigma / \sigma$ Statistical} & {\bf $\Delta\sigma / \sigma$ Systematic} & {\bf $\Delta\sigma / \sigma$ Total} \\
      \hline
      Cut-Based & 1\FB   & 5.0\% & 45\%  & 45\% \\
      BDT       & 1\FB   & 5.7\% & 22\%  & 23\% \\
      Cut-Based & 10\FB  & 1.6\% & 22\%  & 22\% \\
      BDT       & 10\FB  & 1.8\% & 10\%  & 10\% \\
      \hline

    \end{tabular}
  \end{center}
  \caption{Uncertainties on the t-channel cross section measurement for 1\FB\ and 10\FB\ of integrated  luminosity.}
  \label{tbl:t-channel:errors}
\end{table}
%
The single-top cross section is proportional to $|f_{L}V_{tb}|^2$, where $f_L$ is
the weak left-handed coupling and equal to 1 in the Standard Model.  The estimated
uncertainty on $|V_{tb}|$ was calculated to be
${\Delta\left|V_{tb}\right|}/{\left|V_{tb}\right|} = \pm 11.2\%_{stat+sys} \pm 3.9\%_{theor} = \pm 11.9\%.$

The s-channel cut-based analysis was performed by requiring exactly two jets to account for 
event topology, which rejected mostly \ttbar\ events.  The two jets were required to
be $b$-tagged in order to reject $W$ + jets and QCD background, both of which are characterized
by soft $b$-jets or no $b$-jets at all.  In addition, the opening angle between the two $b$-jets
was required be between 0.5 and 4, the scalar sum of the total jet $\mypt$ was required be between $80\mygev$ and
$220\mygev$, and the $\myetmiss$ plus lepton $\mypt$ was required be less than $130\mygev$.  For a 1\FB\ sample,
the cut-based analysis yielded 25 signal events and 251 background events.
The high background levels remaining after the cut-based analysis motivated the use of likelihood
functions (LFs).  Separate LFs were used to discriminate against \ttbar\ events in the
$l + \tau$, $l$ + jets, and di-lepton decay modes, $W$ + jets, and t-channel events.  Input variables 
to the LFs were chosen according to their discrimination power and thresholds were set by 
minimizing the total uncertainty on the cross section.  
For a 1\FB\ sample, there were 15.4 s-channel signal events and 82.7 background events.
The uncertainties are shown in Table~\ref{tbl:s-channel:errors}, where it can be seen that the s-channel cross section
measurement is both statistically and systematically limited.
The dominant systematics on the s-channel cross section measurement were ISR and FSR,
the background cross sections, and luminosity.  
\begin{table}[ht]
  \begin{center}
    \begin{tabular}{|c|c|c|c|c|}
      
      \hline
      {\bf Selection} & {\bf Luminosity} & {\bf $\Delta\sigma / \sigma$ Statistical} & {\bf $\Delta\sigma / \sigma$ Systematic} & {\bf $\Delta\sigma / \sigma$ Total} \\
      \hline
      LF       & 1\FB   & 64\%  & 95\%  & 115\% \\
      LF       & 10\FB  & 20\%  & 48\%  & 52\% \\
      \hline

    \end{tabular}
  \end{center}
  \caption{Uncertainties on the s-channel cross section measurement for 1\FB\ and 10\FB\ of integrated luminosity.}
  \label{tbl:s-channel:errors}
\end{table}

The $Wt$-channel cut-based analysis required one $b$-tagged jet with $\mypt > 50\mygev$ to account
for event topology.  In addition, events having additional $b$-tagged jets
with $\mypt > 35\mygev$ were vetoed in order to reject \ttbar\ events.  
Compared to the b-jets upon which the $\mypt$ requirement was imposed, 
the b-jets used in the veto were selected with a looser b-tag weight
which was optimized according to the signal over
\ttbar\ background ratio.  
The $Wt$-channel cross section was analyzed separately for the different jet multiplicities.
For a 1\FB\ sample, there were 435 signal and 6359 background events for a jet multiplicity of two,
164 signal events and 1088 background events for a jet multiplicity of three, and 40 signal 
and 377 background events for a jet multiplicity of four.
BDTs were used to discriminate the $Wt$-channel signal against \ttbar\ events in the $l$ + jets
and di-lepton channels, $W$ + jets, and t-channel events.  BDT thresholds were set by
minimizing the total uncertainty.  
For a 1\FB\ sample, there were
58 signal and 166 background events for a jet multiplicity of two,
21 signal and 45 background events for a jet multiplicity of three, and 6.6 signal 
and 15.6 background events for a jet multiplicity of four.
Uncertainties are shown in Table~\ref{tbl:Wt-channel:errors}, where it can be seen that
systematics dominate the measurement.
The dominant systematics on the $Wt$-channel uncertainties were ISR and FSR, background
cross sections, and luminosity.
\begin{table}[ht]
  \begin{center}
    \begin{tabular}{|c|c|c|c|c|}
      
      \hline
      {\bf Selection} & {\bf Luminosity} & {\bf $\Delta\sigma / \sigma$ Statistical} & {\bf $\Delta\sigma / \sigma$ Systematic} & {\bf $\Delta\sigma / \sigma$ Total} \\
      \hline
      BDT       & 1\FB   & 21\%   & 48\%  & 52\% \\
      BDT       & 10\FB  & 6.6\%  & 19\%  & 20\% \\
      \hline

    \end{tabular}
  \end{center}
  \caption{Uncertainties on the $Wt$-channel cross section measurement for 1\FB\ and 10\FB\ of integrated luminosity.}
  \label{tbl:Wt-channel:errors}
\end{table}

\section{Conclusions}
For a 5$\sigma$ single-top quark discovery in the t-channel, 
1\FB\ of integrated luminosity is needed while 30 \FB\ is needed for 3$\sigma$ evidence
in the s-channel.  1 \FB\ is needed for evidence and 10 \FB\ is needed for discovery in the $Wt$-channel.
Systematic uncertainties are the limiting factor in the single-top cross section measurement, except for the
s-channel where statistics also play a limiting role.  The systematics in the current analysis have a strong 
dependence on Monte Carlo and the use of data-driven analysis techniques should help minimize them.


\end{document}